\documentclass[aps,prb,twocolumn,superscriptaddress, show pacs]{revtex4-1}
\usepackage{bm, amsmath}
\usepackage{graphicx}
\usepackage{color}
\usepackage{epstopdf}
 
\begin{document} 

\title{Discrete helicoidal states in chiral magnetic thin films}

\author{M. N. Wilson}
\author{E. A. Karhu}
\author{D. P. Lake}
\author{A. S.  Quigley}
\author{S. Meynell}
\affiliation{Department of Physics and Atmospheric Science, Dalhousie University, Halifax, Nova Scotia, Canada B3H 3J5}

\author{A.~N.~Bogdanov}
\affiliation{IFW Dresden, Postfach 270116, D-01171 Dresden, Germany}

\author{H. Fritzsche}
\affiliation{Atomic Energy of Canada Limited, Chalk River, Ontario, Canada K0J IJ0}

\author{U.~K.~R\"{o}{\ss}ler}
\affiliation{IFW Dresden, Postfach 270116, D-01171 Dresden, Germany}

\author{T.~L.~Monchesky}
\affiliation{Department of Physics and Atmospheric Science, Dalhousie University, Halifax, Nova Scotia, Canada B3H 3J5}

\email[]{theodore.monchesky@dal.ca}
\thanks{}

\date{\today}

\begin{abstract}
Magnetoresistance (MR), polarized neutron reflectometry (PNR) and magnetometry measurements in
MnSi thin films and rigorous analytical solutions
of the micromagnetic equations show that the field-induced unwinding of confined helicoids
occurs via discrete steps.
A comparison between the theoretical results and the PNR and magnetometry data
shows that finite size effects confine the wavelength 
and lead to a quantization of the number of turns in the helicoid.  
We demonstrate that the magnetic state of these finite helicoids can be read by electrical means.
\end{abstract}


\maketitle

\section{Introduction}
The Dzyaloshinskii-Moriya (DM) interaction \cite{Dzyaloshinskii:1964jetp} and  epitaxial induced strain in thin film cubic helimagnets \cite{Karhu:2010prb,Karhu:2011prb,Karhu:2012prb,Huang:2012prl, Porter:2012prb}  induce complex magnetic textures over a wide range of temperatures.\cite{Karhu:2012prb, Huang:2012prl, Wilson:2012prb}  Particularly, chiral skyrmions,\cite{Bogdanov:1994jmmm,Rossler:2011jpcs} usually metastable in bulk cubic helimagnets, are predicted to be thermodynamically stable in nanolayers of these materials.\cite{Karhu:2012prb,Butenko:2010prb,Rybakov:2013prb}

Recently, these specific solitonic states have been observed in thin films\cite{Huang:2012prl, Wilson:2012prb} and mechanically thinned nanolayers\cite{Yu:2010nat,Tonomura:2012nl}  over extended regions of the magnetic phase diagram. These states are freely created and driven with an ultra-low current density, and are considered as promising objects for novel types of magnetic data storage.\cite{Kiselev:2011jpd,Yu:2012nc} The induced anisotropy also extends the region of the magnetic phase diagram occupied by distorted helical states (\textit{helicoids})\cite{Butenko:2010prb, Karhu:2012prb}, which have been proposed for a number of spintronic applications.\cite{Kishine:2011prl, Bostrem:2008prb-a, Borisov:2009prb}

Bulk MnSi belongs to the cubic $P2_13$ ($T^4$) space group of magnetic crystals that lack inversion symmetry, which through spin-orbit coupling produces the DM interaction responsible for the long wavelength helical order.\cite{Dzyaloshinskii:1964jetp,Bak:1980jpc}
Owing to a small cubic anisotropy, the helical order does not unwind in a magnetic field since modest fields cause a reorientation of the helix along the field direction.\cite{Plumer:1981jpc}  In bulk crystals, a field-induced ferromagnetic state is created by a second-order transition via a conical phase, as observed in MnSi,\cite{Okorokov:2005sj} FeGe,\cite{Lebech:1989jpcm} and Cu$_2$OSeO$_3$.\cite{Seki:2012sci}  However, in lower symmetry crystals, the DM interaction can be uniaxial (e.g. Cr$_{1/3}$NbS$_2$ \cite{Kousaka:2009ni, Togawa:2012prl}).
In this case, 
a field that is applied transverse to the helical propagation vector does not reorient the helix, 
but rather it creates helicoidal distortions.  While bulk MnSi belongs to the former group, the strain induced by the Si substrate in epitaxial MnSi thin films lowers the symmetry to a trigonal $R3$ ($C_3^4$) space group and creates 
 a uniaxial anisotropy with the hard-axis along the film normal. 
This anisotropy helps to stabilize the helicoidal state with an in-plane magnetization and the propagation vector perpendicular to the film surface.\cite{ Karhu:2010prb, Karhu:2011prb}
A transverse field unwinds the helices and transforms them via first-order transitions into either elliptic skyrmion or elliptic cone phases.\cite{Karhu:2012prb, Wilson:2012prb}

This Article investigates the role of finite size in the confinement of helicoids in MnSi thin films.  Unlike bulk crystals, we observe discrete jumps in the magnetization at specific fields and thicknesses that correspond to the annihilation of individual turns of the helicoid.
These jumps result from the truncation of the helicoids by the film interfaces, which stabilizes a quantized number of turns in the helicoids.
The sudden changes in magnetic structure during the unwinding of the helicoids in a magnetic field produce anomalies in
the magnetoresistance (MR) that enable an electronic reading of the number of turns in the helicoidal structure. 

\section{Theory}
\label{sec:theory}
Chiral modulations in low-anisotropy helimagnets are well described theoretically 
by the standard Dzyaloshinskii model.\cite{Dzyaloshinskii:1964jetp, Bak:1980jpc, Karhu:2012prb}
We define $\mathbf{m} = \mathbf{M}/M_s$ as a unit vector in the direction of the magnetization $\mathbf{M}$ ($M_s = |\mathbf{M}|$), and $\mathbf{H}$ as the field applied along the $x$-axis.
For a helicoid propagating along the $z$-axis, $\mathbf{m} = ( \sin \theta, \cos \theta,0)$ and
the energy density in the Dzyaloshinskii model is reduced to 
\begin{equation}
\label{eq:w}
w (\theta) =  A \, \left( \frac{d\theta}{dz} \right)^2 - D \,\frac{d\theta}{dz}- HM_s \cos \theta,
\end{equation}
in terms of the exchange stiffness energy with constant, $A$, the Dzyaloshinskii-Moriya coupling with constant, $D$, and the Zeeman energy.  Model (\ref{eq:w}) introduces two fundamental parameters: 
the zero field \textit{helical wavelength}, $L_D = 4 \pi A/D$ and the
\textit{saturation field} 
in the zero anisotropy limit, 
$H_D = D^2 /(2 A M_s)$.  
In the case of MnSi films, $L_D=13.9$~nm,\cite{Karhu:2011prb} and the average $H_D$  is 0.77~T  with a standard deviation of 0.05 T for the range of sample thicknesses presented in this Article. 
 These parameters are tabulated for a number of other cubic helimagnets in  Refs.~\onlinecite{Rossler:2011jpcs, Wilhelm:2012jpcm}. 

Minimization of functional $w (\theta)$
leads to the well known differential equation 
for a non-linear pendulum.\cite{Dzyaloshinskii:1964jetp}
Analytical solutions for helicoids in bulk helimagnets with uniaxial anisotropy predict a continuous transformation
from a single-harmonic helix with period $L(0)=L_D$ at zero field
into a set of isolated kinks at critical field  
$H_h / H_D  = \pi^2/ 16 = 0.618$,\cite{Dzyaloshinskii:1964jetp} 
which agrees well with observations in 
Ref.~\onlinecite{Togawa:2012prl}. %

The unique magnetic properties of magnetic nanostructures are generally a result of reduced dimensionality and surface/interface induced magnetic anisotropy.\cite{Gradmann:1986jmmm, Bruno:1989apa, Heinrich:1993ap} For confined helicoids, these two factors can be readily taken into account by minimization of functional (\ref{eq:w}) in a layer of finite thickness with boundary conditions including surface/interface magnetic anisotropy.  Our previous work did not find any significant influence of interface anisotropy on the magnetic properties of the thin films in the range of thicknesses presented here.\cite{Karhu:2011prb} However, the breaking of translational symmetry in finite helicoids due to the presence of interfaces substantially influences their properties. For an in-plane magnetic field, confined helicoids in nanolayers evolve into the field induced ferromagnetic phase via a number of discrete jumps between states that have a quantized number of helicoidal turns. We calculate these states by minimization of functional (\ref{eq:w}) in a layer of finite thickness $d$ with \textit{free} boundary conditions. We have derived analytical solutions for confined helicoids in the Appendix and present their main features in Fig.~\ref{fig:model}.

\begin{figure}[h]
\centering
\includegraphics[width=8 cm]{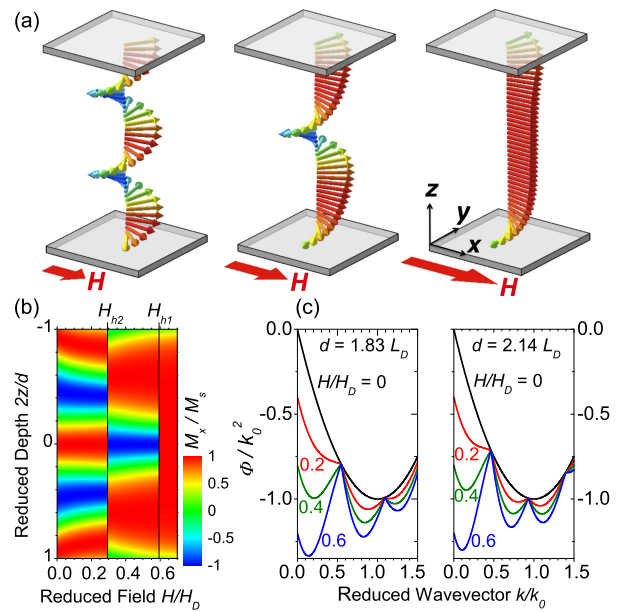}
\caption{(color on-line) 
(a) Calculated spin configurations for a helimagnet thin film of thickness $d = 2.14 L_D$ with applied in-plane fields $H=0.25H_D$ (left), $0.55H_D$ (middle), $0.61 H_D$ (right).
(b) Component of $\mathbf{M}$ along $\mathbf{H}$ as a function of field and depth in the film of thickness $d=2.14 L_D$.
(c) The average energy density for the linear approximation given by Eq.(2) for $d = 1.32 L_D$ (left), and $d = 2.14 L_D$ (right) in applied fields $H = 0.0, 0.2, 0.4,$ and 0.6 $H_D$
}
\label{fig:model}
\end{figure}

Figure~\ref{fig:model}(a) shows the three quantized states that are supported in a film with thickness $d = 2.14 L_D$  with either zero, one, or two turns of the helix, depending on the applied field. The details of the evolution of the magnetic structure of this film as a function of applied field is presented in Fig.~\ref{fig:model}(b).  The figure shows that the wavelength varies weakly with applied field up to $H_{h1} = 0.28 H_D$. Above  $H_{h1}$, one turn of the helicoid is pushed out, and the second turn is pushed out above $H_{h2} = 0.60 H_D$.  For $H > H_{h2}$, the system is in a twisted ferromagnetic state, with a ferromagnetic state in the center of the film and chiral modulations at each interface. Our theory predicts that this state will persist up to high field.
Alternatively, depending on thickness and anisotropy, the system could transition from a single turn state to a conical phase with an in-plane propagation vector.\cite{Rybakov:PC}

To elucidate the remarkable features of the quantized helicoid structure, we consider helicoids described by a simplified linear function $ \theta (z) = k z + \theta_0 $, and ignore the higher harmonics that only become significant near the critical field $H_{h}=0.618$.\cite{Izyumov:1984spu}  The Zeeman energy reaches a minimum for the largest possible net magnetization along the applied field, which implies that the  moments in the center of the film are either parallel ($\theta(0) =0$) or antiparallel  ($\theta(0) =\pi$) to the applied field depending on the number of turns of the helix (cf. Fig. 1 (b)). By inserting these values of $\theta_0$ into the average energy density $\bar{w} = (1/d) \int_{-d/2}^{d/2} w(\theta) dz$, we obtain the following expression for the reduced energy density $\Phi = \bar{w}/ A$ as a function of wavevector k,
\begin{eqnarray}
\label{eq:w2}  
\frac{\Phi (k)}{k_0^2}  = \left(\frac{k}{k_0}\right)^2 -  \frac{2 k}{k_0} - \frac{ 4 }{k d}
\left| \sin\!\left( \frac{k d}{2}\right)\! \right|  \frac{H}{H_D}, 
\end{eqnarray}
where $k_0 = 2\pi/L_D$ is the wavevector in zero field.
We plot $\Phi(k)$ for thicknesses $d = 1.83 L_D$ and $2.14 L_D$ for different
values of the applied field in Fig.~\ref{fig:model}(c) to show that a potential well corresponding to the undisturbed helix $k = k_0$ splits into two wells at finite fields, one with a minimum at $k_{+1} > k_0$, and  the other with a minimum at  $k_{-1} < k_0$.   At higher fields, additional local minima form. These wells arise due to $k$-dependent oscillations in the Zeeman energy from the uncompensated moments in the finite helicoids. Transitions between helicoid states with a different number of turns requires the system to jump between these energy minima; the energy barrier inherent in this process causes these transitions to be first order.

One of the features of helicoidal films that distinguishes them from bulk behaviour is their evolution in small field. The low field slope of the helicoid wavelength in the low field limit, $\eta_0 = (H_D/L_D)(d L/dH)_{H =0}$,
can be readily derived as a function of the film thickness by an expansion of the potential $\Phi$ in Eq.~\ref{eq:w2} for small values of $H$:
  
\begin{eqnarray}
\label{eq:slope}  
\eta_0 (\nu)  = \frac{1}{4} \left(  \frac{\sin \pi \nu}{\pi \nu}  -\cos \pi \nu \right) 
\mathrm{sgn} \left( \sin \pi \nu \right).
\end{eqnarray}
In this equation, the \textit{confinement ratio}, 
$ \nu = d/L_D$, determines the sign of $\eta_0$.  Surprisingly, $\eta_0$ can take on negative values that correspond to the helicoid tightening its pitch with increasing field as the system evolves in the $k _{+1}$ potential well.  The predicted tightening of $L$ for the $d=2.14 L_D$ film  is visible in Fig.~\ref{fig:model}(b).  For positive $\eta_0$, the pitch relaxes with applied field because the helicoid resides in the $k_{-1}$ potential well, as illustrated by the calculation for $d=1.83 L_D$ in Fig.~\ref{fig:model}(c).  The thickness dependence of the oscillations in $\eta_0$ are presented in Fig.~\ref{fig:MR}(c).
After an initial evolution in either the $k _{+1}$ or $k_{-1}$ wells, rigorous solutions for Eq.~\ref{eq:w} show that the helicoids transition via a series of first-order transitions toward the lowest $k$-state that corresponds to the twisted ferromagnetic state shown on the right in Fig.~\ref{fig:model}(a).

\section{Experiment}
The epitaxial MnSi thin films studied in this Article were grown on high resistivity Si wafers by molecular beam epitaxy, as detailed in Ref.~\onlinecite{Karhu:2011prb}. All measurements on these samples were collected with the magnetic field applied along the MnSi [1$\overline{1}$0] direction.

\subsection{Magnetoresistance}
The predicted oscillations in $\eta_0(\nu)$ have important implications for the magnetoresistance (MR),  defined by  $\Delta \rho / \rho_0 = (\rho(H) - \rho(0) )/\rho(0)$.  
Resistivity measurements were performed with a four-probe setup using gold wires soldered onto the films with indium solder. The samples used for these measurements were either cleaved strips, or Hall bar arrangements that were prepared by photolithographic patterning using SPR220 3.0 photoresist and a hydrofluoric/nitric/acetic acid etch. No systematic differences were observed in the resistivity measurements of the samples prepared by these two methods.

\begin{figure}[h]
\centering
\includegraphics[width=8 cm]{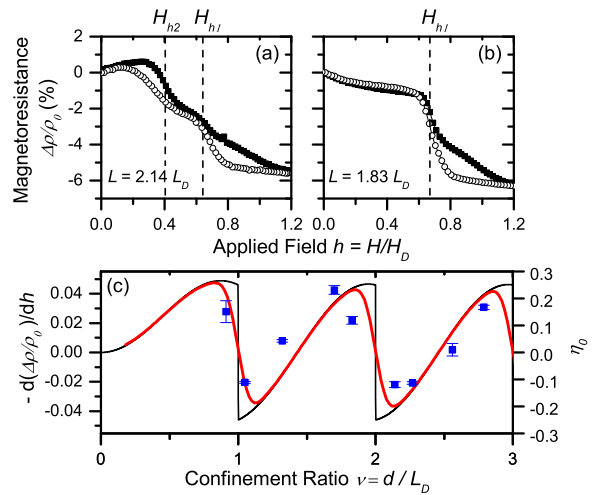}
\caption{Magnetoresistance measured at $T=5$~K for film thicknesses 2.14$L_D$ (a) and 1.83$L_D$ (b) in a field $\mathbf{H} \| \textrm{MnSi}[1\overline{1}0]$. Open circles are the decreasing field data, closed squares are increasing field. Marked fields $H_{h1}$ and $H_{h1}$ are the transition fields taken from magnetometry measurements from the increasing branch data.  (c) The slope of the magnetoresistance (blue squares) compared to $\eta_0$ (black line) from Eq.~(\ref{eq:slope}).  The red line shows $\eta_0$ with 0.9 nm rms thickness variations.
}
\label{fig:MR}
\end{figure}

Measurements in Fig. \ref{fig:MR} show that for the $d=2.14 L_D$ film with $\eta_0 < 0$, there is an anomalous positive magnetoresistance in low fields, in contrast to the $\eta_0 > 0$ films that show a more conventional  $\Delta \rho < 0$, as demonstrated by the $d=1.83 L_D$ film measurements in Fig.~\ref{fig:MR}(b).  This can be explained by an expected increase in MR with decreasing $L$.\cite{Viret:1996prb,Levy:1997prl}  Therefore, in Fig.~\ref{fig:MR}(c) we compare the thickness dependence of $d \Delta \rho / d H$ measured at $H=0$ to $\eta_0$ and find that the oscillations in the slope of $\Delta \rho$ follow the predicted oscillations in the slope of the wavelength.  To account for inhomogeneities in thickness, we convolve $\eta_0$ (thin black line) with a gaussian of width specified by the typical rms roughnesses of 0.4~nm and 0.8~nm for the top and bottom interfaces, as determined by x-ray reflectometry.  The result is shown by the thick red line.

The magnetoresistance also shows distinct drops as each film approaches the critical fields that signal the forcing out of a turn in the helicoid, labeled $H_{h1}$ and $H_{h2}$ in Fig.~\ref{fig:MR}(a) and (b).  For thicknesses $L_D < d < 2L_D$ we see a single drop in the MR at $H_{h1}$ (Fig.~\ref{fig:MR}(b)) whereas at thicknesses $2L_D < d < 2.5 L_D$ two first-order transitions occur as illustrated by the two well defined hysteretic drops in the MR of Fig.~\ref{fig:MR}(a).

\subsection{Polarized Neutron Reflectometry}
The discrete helicoidal states can be more directly observed using Polarized Neutron Reflectometry (PNR).  PNR measurements were performed on a 26.7-nm MnSi film with the D3 reflectometer at the Canadian Neutron Beam Centre equipped with the M5 superconducting magnet cryostat. This instrument uses 0.237~nm neutrons that are polarized using an Fe/Si supermirror and a Mezei-type precession spin flipper to achieve a typical spin polarization exceeding 90\%. All PNR measurements were collected by field cooling the sample from above $T_C$ to a temperature of $T=5$~K in a field of 2~T, and measuring the reflectivity after decreasing the field to the desired value.  Only the non-spin-flip cross-sections were measured since previous measurements show that the components of the magnetization perpendicular to the field cancel due to the presence of chiral domains.\cite{Karhu:2012prb}

\begin{figure}[]
\centering
\includegraphics[width= 8 cm]{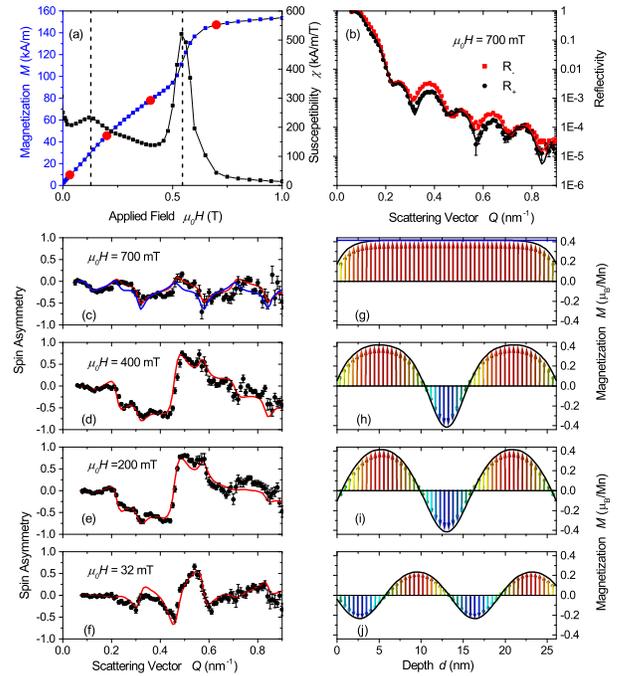}
\caption{(a) Magnetization (blue squares) and magnetic susceptibility (black squares) for the $1.92L_D$ sample at $T = 5$~K, plotted with the PNR measurement fields indicated by the red circles, and transition fields shown by dotted lines. (b) PNR cross-section for $R_{-}$ (red squares) and $R_{+}$ (black circles) reflectivities for the $1.92L_D$ sample measured at $T = 5$~K, $\mu_0 H = 0.7$~T.   Solid lines show the calculated spin-up (red) and spin-down (black) reflectivity. (c)-(f) Measured (black circles) and calculated (red line) spin asymmetry for the $1.92L_D$ sample at $T = 5$~K, $\mu_0 H = $700, 400, 200, and 32 mT. (g)-(j) Magnetization depth profiles used to calculate the spin asymmetries shown in figures (c)-(f). The blue line in (c) shows the spin asymmetry calculated for the ferromagnetic state with a depth profile shown by the blue line in (g).  All error bars are $\pm 1\sigma$.
} 
\label{fig:PNR}
\end{figure}

 We measured PNR spectra at field values corresponding to the red circles in Fig.~\ref{fig:PNR} (a), which were chosen to determine the magnetic states on either side of the two transition fields, indicated by the dashed lines. One such PNR data set is shown by the red and black points in Fig.~\ref{fig:PNR} (b) for $T = 5$~K and $\mu_0 H = 0.7$~T.  We compare the PNR data directly to theory without any fitting parameters. The nuclear scattering length densities are derived from x-ray reflectometry measurements, whereas the magnetic scattering length densities are calculated directly from the discrete helicoidal model described in Section~\ref{sec:theory} with $L_D=13.9$~nm,\cite{Karhu:2011prb} $H_D=0.82 T$, and $M_s = 0.415 \mu_B$/Mn.\cite{Karhu:2012prb}   Reflectivity curves calculated using the Simulreflec software package are in excellent agreement with the data (Figure~\ref{fig:PNR}(b)).

In order to clearly separate the nuclear scattering from the magnetic scattering, we plot the spin asymmetry, $(R_{+} - R_{-})/(R_{+} + R_{-})$ in Figs.~\ref{fig:PNR} (c)-(f). Next to each of these figures we plot the calculated the field-dependent magnetization depth profiles in Figs.~\ref{fig:PNR} (g)-(i). These are obtained by minimizing the functional of Eq.~(\ref{eq:w}), and are used to calculate the red curves in Figs.~\ref{fig:PNR}(c)-(e) without any fitting parameters.  Figures~\ref{fig:PNR} (c) and (g) demonstrate that the PNR data at 700~mT are consistent with the calculated twisted ferromagnetic state (red line), and are inconsistent with a cone phase with an in-plane propagation vector or a pure ferromagnetic state (blue line), as is particularly evident for the region 0.2 nm$^{-1} <$ Q $<$ 0.3 nm$^{-1}$.  Below the field $H_{h1}$, Fig.~\ref{fig:PNR}(d) and (h) show the nucleation of a solitonic kink in the magnetization profile, which relaxes and becomes more sinusoidal in Fig.~\ref{fig:PNR}(e) and (i). The fitting parameter-free agreement between the calculated depth profiles and the PNR data at these three fields gives strong evidence that our theory accurately predicts the magnetic states of this film.

For the lowest field value in Fig.~\ref{fig:PNR}(f) and (j), a more complex arrangement of the magnetic domain structure arises due to the frustration between crystal domains with opposite chirality.\cite{Karhu:2010prb}  In this case, the simple model given by Eq.~(\ref{eq:w}) is insufficient to predict the behaviour.  A fit to the data in Fig.~\ref{fig:PNR}(f) shows a reduction of the average magnetization and a phase shift of the helix represented by the depth profile in Fig.~\ref{fig:PNR}(j). 

Our previous PNR measurements showed unexplained behavior at $\mu_0 H = 0.5$~T, where the depth profile could not be fit by a single helicoid.\cite{Karhu:2012prb}  However, these new results are now explained with the discrete helicoid model.  The peak in the static susceptibility at $\mu_0 H_{h1} = 0.54$~T represents a first-order magnetic phase transition.  Such transitions are characterized by the coexistence of magnetic phases, and a superposition of the magnetic profiles on either side of the transition are able to fit the magnetic scattering length density of the $\mu_0 H = 0.5$~T data in Fig. 9 (c) of Ref.~\onlinecite{Karhu:2012prb}.  

\subsection{Magnetometry}

\begin{figure}[h]
\centering
\includegraphics[width=7.5 cm]{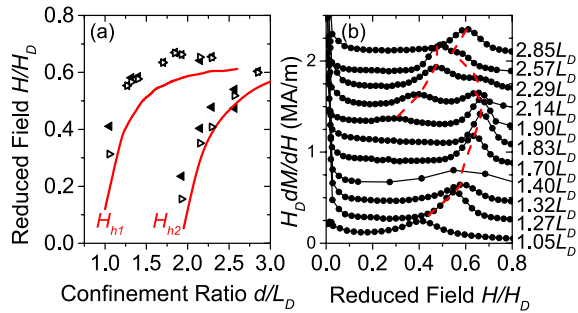}
\caption{(color on-line) 
(a) Critical fields extracted for increasing (filled triangles) and decreasing (open triangles) magnetic fields with $\mathbf{H} \| \textrm{MnSi}[1\overline{1}0]$ at $T=5$~K.  The solid red lines in (a) are the calculated threshold fields for the removal of a turn in the helicoid. (b) DC susceptibility measurements in an increasing magnetic field for thicknesses $L_D < d <  2.85 L_D$, offset vertically for clarity.  Red dotted lines track the observed transition fields.
} 
\label{fig:HvD}
\end{figure}

As further evidence for these discrete helicoid states, we have analyzed SQUID magnetometry data measured on samples of thickness $L_D < d < 3L_D$. 
Magnetometry data was collected using a Quantum Design MPMS-XL-5 SQUID magnetometer operating in the reciprocating sample option on cleaved 4~mm $[1\overline{1}0]$ x 6~mm sections of the MnSi films.
The peaks in the field dependence of the static susceptibility $dM/dH$ in Fig.~\ref{fig:HvD}(b) confirm that there are either one or two first-order magnetic phase transitions below the saturation field $H_{sat}$.  The transition fields obtained from these peaks are plotted in Fig.~\ref{fig:HvD}(a) after being normalized by $H_D$, which is estimated from the field $H_{sat}$ indicated by a minimum in $d^2M/dH^2$.  The analysis used to relate $H_D$ to the saturation field is  described in Ref.~\onlinecite{Karhu:2012prb}. 

Figure~\ref{fig:HvD}(a) compares the measured $H_{h1}$ and $H_{h2}$ to the theoretical threshold fields where a single turn of the helicoid is annihilated (red lines).  The observed transitions occur at fields that are approximately $0.05H_D$ higher than predicted. These discrepancies may be due to interfacial anisotropies and/or the softening of the exchange and Dzyaloshinskii-Moriya interactions at the interfaces that are not accounted for in the model. The slight field discrepancy may also be a result of a systematic error in measurement of $H_{sat}$ used to calculate $H_D$, since the minimum in $d^2M/dH^2$ is weak and broad at $T = 5$~K.  The calculation of $H_D$ is further complicated by the twisting of the magnetization at the interfaces shown in the twisted ferromagnetic state in Fig.~\ref{fig:model}(a), as the procedure we use to estimate $H_D$ assumes a conical state at high field. Further work will be needed to address the effect this has on the estimation of $H_D$. 

\section{Conclusions}

In conclusion, we have shown that a confining geometry in helimagnetic films with a strong easy-plane anisotropy stabilizes a quantized number of turns in the helicoid. The magnetization processes in the helicoidal state display the required functionalities of a device based on chirally twisted states, namely, (i) the states are discrete and reproducible, (ii) switching between these states is in principle possible due to the metastability demonstrated by the observed hysteresis, and (iii) the state of the helicoid can be read by electronic means.  Furthermore, in higher magnetic fields, the thicker films are able to confine solitonic kinks that open the possibility to use cubic helical magnetics to explore some of the predicted effects in a helicoid soliton lattice.\cite{Kishine:2011prl, Bostrem:2008prb-a, Borisov:2009prb}

\begin{acknowledgements}
We thank F. N. Rybakov for fruitful discussions and M. Johnson for technical assistance.  TLM and MNW acknowledge support from NSERC, and the support of the Canada Foundation for Innovation, the Atlantic Innovation Fund, and other partners which fund the Facilities for Materials Characterization, managed by the Institute for Research in Materials. The research presented herein is made possible by a reflectometer jointly funded by Canada Foundation for Innovation (CFI), Ontario Innovation Trust (OIT), Ontario Research Fund (ORF), and the National Research Council Canada (NRC)
\end{acknowledgements}

\appendix*
\section{}
Here we derive analytical solutions for helicoids described by
Eq.~(\ref{eq:w}) for a layer of thickness $d$ and free boundary conditions.
The Euler equation for functional (\ref{eq:w}) has the first
integral 

\begin{eqnarray}
\frac{1}{k_0^{2}} \left( \frac{d \theta}{d  z} \right)^2 + h \cos \theta = c,
\end{eqnarray}
where $c$ is an integration constant, and $h$ is the reduced field $H/H_D$ and $k_0 = D / (2A) = 2 \pi / L_D$ is the propagation vector in zero field.\cite{Dzyaloshinskii:1964jetp}  The film interfaces break the translational symmetry of the helicoids, and for free boundary conditions at finite magnetic fields,  Eq.~(\ref{eq:w}) has solutions with either $\theta (0) = 0$ or $\pi$. The solution to the Euler equations is given by, 
\begin{eqnarray}
\label{eq:solution1}  
k_0 z =\int_{0}^{\theta} \frac{d t}{\sqrt{c -2 h \cos t}}.
\end{eqnarray}
For $h > 0$,  Eq.~(\ref{eq:solution1}) describes 
magnetization profiles with $\theta_0 = \theta(0) =0$, and for $h <0$, profiles
with $\theta_0 = \pi$.
%
\begin{figure}[!b]
\centering
\includegraphics[width=6 cm]{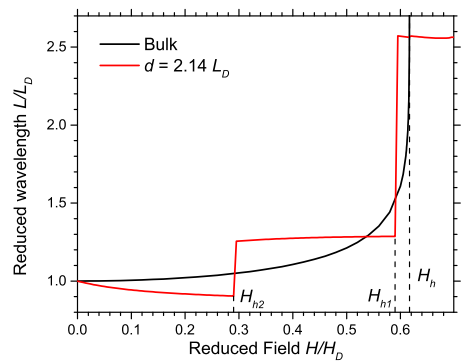}
\caption{(color on-line) The field dependence of the wavelength for a bulk crystal (black), and a film with thickness $ d= 2.14 L_D$ (red).  The dashed lined indicates the critical fields $H_{h2}$ and $H_{h1}$, as well as $H_h = \pi^2 H_D /16$ where an infinite helicoid breaks into a series of isolated $360^\circ$ domain walls.}
\label{fig:L}
\end{figure}
Equation~(\ref{eq:solution1}) can be expressed in terms of an incomplete elliptic integral of the first kind, and then inverted in order to find an expression for $\theta$ in terms of the Jacobi amplitude function, 
\begin{eqnarray}
 \theta(z) = 2 \ \textrm{am}\!\left( k_0 z \sqrt{h}/\kappa ; i \kappa \right) + \theta_0
 \label{eq:theta}
 \end{eqnarray}
 where $\kappa$ is the modulus of the elliptic function:\cite{Abramowitz:1965}
\begin{eqnarray}
\kappa  = \sqrt{\frac{4 h }{c - 2 h}}.
\label{eq:mod}
\end{eqnarray}
The equilibrium helicoid configuration  $\theta(z)$ is a function of the two control parameters $d$ and $h$, and is obtained by finding $\kappa$ that minimizes the energy density (\ref{eq:w}) averaged over the layer thickness,
\begin{eqnarray}
\label{eq:wbar}  
\bar{w}(\kappa) = \frac{1}{d} \int_{-d/2}^{d/2} w\left( 2  \ \textrm{am}\!\left( k_0 z \sqrt{h}/\kappa ; i \kappa \right) \right) dz.
\end{eqnarray}

The modulus obtained from the minimization of Eq.~(\ref{eq:wbar}) is related to the wavelength, $L$, which is derived from the condition $\theta(z=L/2)=\pi/2$:\cite{Dzyaloshinskii:1964jetp}
 \begin{eqnarray}
 \frac{L(h)}{L_D} = \frac{\kappa}{\pi \sqrt{h}}K(\kappa).
 \label{eq:L}
 \end{eqnarray}
The difference between the solution for a bulk crystal and that of a thin film is illustrated in Fig.~(\ref{fig:L}), which shows the anomalous tightening of the helicoid below $H_{h2}$ and the discontinuities in the wavelength at fields $H_{h1}$ and $H_{h2}$.


%

\end{document}